\documentclass[prd,eqsecnum,twocolumn,showpacs,amsmath,amssymb]{revtex4}

\usepackage{graphicx}

\usepackage{bm}

\newcommand{\beq}{\begin{equation}}
\newcommand{\eeq}{\end{equation}}
\newcommand{\beqs}{\begin{eqnarray}}
\newcommand{\eeqs}{\end{eqnarray}}

\begin{document}

\title{Exact Results on Potts/Tutte Polynomials for Families of Networks with 
Edge and Vertex Inflations} 

\author{Robert Shrock}

\affiliation{ C. N. Yang Institute for Theoretical Physics \\
Stony Brook University \\
Stony Brook, NY 11794}

\begin{abstract}

We derive exact relations between the Potts model partition function, or
equivalently the Tutte polynomial, for a network (graph) $G$ and a network
obtained from $G$ by (i) by replacing each edge (i.e., bond) of $G$ by two or
more edges joining the same vertices, and (ii) by inserting one or more
degree-2 vertices on edges of $G$. These processes are called edge and vertex
inflation, respectively.  The physical effects of these edge and vertex
inflations are discussed.  We also present exact calculations of these
polynomials for families of networks obtained via the operation (ii) on a
subset of the bonds of the network.  Applications of these results include
calculations of some network reliability polynomials. In addition, we evaluate
our results to calculate various quantities of structural interest such as
numbers of spanning trees, etc., and to determine their asymptotic behavior for
large networks.

\end{abstract}

\pacs{05.50.+q,64.60.Cn,75.10.Hk}

\maketitle

\section{Introduction}

The Potts model has long been valuable as a system exhibiting many-body
cooperative phenomena \cite{wurev}. On a lattice, or, more generally, on a
network (i.e., graph) $G$, at temperature $T$, the partition function for this
model (in zero external field) is $Z= \sum_{\{\sigma_i\}}e^{-\beta {\cal H}}$,
where $\beta = 1/(k_BT)$, the Hamiltonian ${\cal H} =
-J\sum_{e_{ij}}\delta_{\sigma_i \sigma_j}$, $J$ is the spin-spin exchange
constant, $i$ and $j$ denote vertices (sites) on $G$, $e_{ij}$ is the edge
(bond) connecting them, and $\sigma_i$ are classical spins taking values in the
set $\{1,...,q\}$. We use the notation $K \equiv \beta J$ and $v \equiv e^K-1$.
Thus, for the Potts ferromagnet (FM, $J > 0$) and antiferromagnet (AFM, $J <
0$), the physical ranges of $v$ are $v \ge 0$ and $-1 \le v \le 0$,
respectively.  In general, a graph $G=(V,E)$ is defined by its vertex set, $V$,
and its edge set, $E$. We denote the number of vertices of $G$ as $n \equiv
n(G) \equiv |V|$ and the number of edges of $G$ as $e(G) \equiv |E|$.  $Z$ is a
polynomial in $q$ and $v$, as will be evident from Eq. (\ref{cluster}) below.
The Potts model partition function $Z$ is equivalent to a function of
considerable interest in modern mathematical graph theory, namely the Tutte
polynomial \cite{tutte}-\cite{boll}, and we shall therefore often refer to this
object in a unified manner as the Potts/Tutte polynomial. The partition
function of the zero-temperature Potts antiferromagnet is identical to another
function of longstanding interest in graph theory, namely the chromatic
polynomial, which counts the number of ways of assigning $q$ colors to the
vertices of $G$ subject to the condition that no two adjacent vertices have the
same color. These are called proper $q$-colorings of $G$.  An important
property of the Potts antiferromagnet is that for sufficiently large $q$, it
exhibits nonzero entropy per site at zero temperature and is thus an exception
to the third law of thermodynamics \cite{chowwu,w}.

An interesting problem involving both statistical mechanics and mathematical
graph theory is to relate the Potts/Tutte polynomial calculated for a graph $G$
with the corresponding polynomial calculated for a graph $\tilde G$ which is
obtained by a specified modification of $G$.  In this paper we will first
present a general solution to this problem for two classes of $\tilde G$'s,
namely those obtained (i) by replacing each edge of $G$ by two or more edges
joining the same vertices, and (ii) by inserting one or more degree-2 vertices
on each edge of $G$ (where the degree, $\kappa_{v_i}$, of a vertex $v_i$ is
defined as the number of edges that connect to it). We denote these operations
as edge and vertex inflations of $G$, respectively.  In the literature on
mathematical graph theory, a vertex inflation of $G$ is also called a
homeomorphic inflation of $G$, and the reverse procedure, of removing one or
more degree-2 vertices from edges of $G$, is called a homeomorphic reduction of
$G$.  In the remaining sections of our paper we will present results for
Potts/Tutte polynomials of certain families of graphs obtained by vertex
inflations for a subset of edges. These families include longitudinal vertex
inflations of free, cyclic, and M\"obius ladder graphs of arbitrarily great
length and a family that we call hammock graphs.  These results extend our
previous work in two ways: (a) as generalizations to the full 
Potts/Tutte polynomials of our earlier calculations with S.-H. Tsai of 
the chromatic polynomials for homeomorphic inflations of these families of
graphs in Refs. \cite{wa3}-\cite{pg} and (b) as homeomorphic inflations of our
previous calculation of the Potts/Tutte polynomials for ladder strips in
Ref. \cite{a}.

There are several motivations for this work. For an arbitrary graph $G$, the
calculation of the Potts/Tutte polynomial involves a number of computational
steps and a corresponding time that grow exponentially rapidly with $n(G)$ and
$e(G)$ (e.g., \cite{welsh,jemrev}).  Furthermore, there is no known exact
closed-form solution for $Z$ for arbitrary $q$ and temperature $T$ on the
thermodynamic limit of lattice graphs of dimensionality $d \ge 2$.  Hence, it
is of fundamental value to carry out exact analytic calculations of Potts/Tutte
polynomials on various families of graphs, such as lattice strip graphs and
modifications thereof.  Furthermore, special cases of the Potts/Tutte
polynomial are of considerable interest in their own right. We have noted the
importance of the zero-temperature Potts antiferromagnet (chromatic
polynomial).  Thus, another motivation for our work, as embodied in point (a)
above, is to generalize our previous calculations with S.-H. Tsai
\cite{wa3}-\cite{pg} to the broader context of the finite-temperature Potts
antiferromagnet and the Potts ferromagnet (the latter also at arbitrary
temperature).  A central motivation is to investigate the effects of edge and
vertex inflations of a graph on the Potts/Tutte polynomial of that graph. This
involves the generalization in point (b) above.  One particular value of our
present work is its demonstration of the usefulness of relating calculations in
the context of the Potts model to corresponding calculations for the Tutte
polynomial and vice versa.  Finally, special cases of our results yield various
quantities of interest such as reliability polynomials, numbers of spanning
trees, etc. for these families of graphs.

\section{Some General Background} 

In this section we briefly discuss some necessary background that will be used
for our calculations.  There is a useful relation that expresses the Potts
model partition function $Z(G,q,v)$ as a sum of contributions from spanning
subgraphs of $G$.  Here, a spanning subgraph $G'=(V,E')$ has the same vertex
set as $G$ and a subset of the edge set $E$, $E' \subseteq E$.  This
relation is \cite{fk}
\beq
Z(G,q,v) = \sum_{G' \subseteq G} q^{k_c(G')}v^{e(G')} \ , 
\label{cluster}
\eeq
where $k_c(G')$ denotes the number of connected components in $G'$. This
formula shows that $Z(G,q,v)$ is a polynomial in $q$ and $v$.  For the Potts
ferromagnet, Eq. (\ref{cluster}) enables one to generalize $q$ from the
non-negative integers to the non-negative real numbers while keeping $Z(G,q,v)$
positive and hence maintaining a Gibbs measure.

The Tutte polynomial of a graph $G$, denoted $T(G,x,y)$, is defined by 
\cite{tutte}-\cite{boll} 
\beq
T(G,x,y) = \sum_{G' \subseteq G} (x-1)^{k_c(G')-k_c(G)}(y-1)^{c(G')} \ , 
\label{t}
\eeq
where $c(G')$ is the number of linearly independent cycles (circuits) in $G'$.
Let us define 
\beq
x = 1 + \frac{q}{v} \ , \quad y = e^K = v+1 \ , 
\label{qvxy}
\eeq
so that $q=(x-1)(y-1)$. The relation between $Z(G,q,v)$ and $T(G,x,y)$ 
follows directly from Eqs. (\ref{cluster}) and (\ref{t}) and is
\beq
Z(G,q,v) = (x-1)^{k_c(G)}(y-1)^{n(G)}T(G,x,y) \ . 
\label{ztrel}
\eeq
With no loss of generality, we will restrict here to connected graphs $G$, so
$k_c(G)=1$.

For a graph $G$, let us denote $G-e$ as the graph obtained by deleting the edge
$e$ and $G/e$ as the graph obtained by deleting the edge $e$ and identifying
the two vertices that were connected by this edge of $G$.  This operation is
called a contraction of $G$ on $e$. From Eq. (\ref{cluster}), it follows that
$Z(G,q,v)$ satisfies the deletion-contraction relation
\beq
Z(G,q,v) = Z(G-e,q,v)+vZ(G/e,q,v) \ .
\label{dcr}
\eeq
An analogous deletion-contraction relation holds for $T(G,x,y)$.

\section{Types of Inflations of a Graph}

Here we discuss in greater detail the edge and vertex inflation of a graph $G$.
As part of our study, we will consider the infinite-length limit of lattice
strip graphs $G$ of finite width, and also the thermodynamic limit of lattice
graphs $G$ of dimensionality $d \ge 2$.  In these cases, we will often use the
notation $\{G\}$ to indicate these limits.  With no loss of generality, we
begin by assuming that $G$ has no multiple edges. We define an edge inflation
of $G$ to be a graph obtained by replacing one or more of the edges of $G$ by
multiple edges joining the same vertices.  Clearly, this leaves the number of
vertices $n(G)$ invariant. In particular, it is natural to define a uniform
$\ell$-fold edge inflation of $G$ as the graph ${\cal E}_\ell(G)$ obtained by
replacing each edge of $G$ by $\ell$ edges joining the same two vertices.
Thus, the number of edges of ${\cal E}_\ell(G)$ is $e({\cal E}_\ell(G))=\ell
e(G)$.  A $\kappa$-regular graph is a graph with the property that all vertices
have the same degree, $\kappa_{v_i} = \kappa \ \forall \ v_i \in V$.  Although
a general graph is not $\kappa$-regular, one can define an average or effective
vertex degree $\kappa_{eff}$ as 
\beq
\kappa_{eff} = \frac{\sum_{v_i \in V} \kappa_{v_i}}{n(G)} \ . 
\label{kappaeff}
\eeq
Clearly, a uniform $\ell$-fold inflation of all of the edges of $G$ multiplies
$\kappa_{eff}$ by $\ell$.  A remark is in order here concerning loops.  A loop
is defined as an edge joining a vertex back to itself.  This would not occur in
the statistical mechanical framework, since it would mean a spin $\sigma_i$
interacting with itself, and hence we will usually assume that $G$ is a
loopless graph.

We define a vertex (i.e., homeomorphic) inflation of $G$ to be a graph obtained
by inserting one or more degree-2 vertices on one or several edges of $G$
\cite{evd}. This leaves the number of (linearly independent) circuits in $G$,
$c(G)$, invariant.  Homeomorphic inflations and reductions of graphs have been
of interest in the study of chromatic and Tutte polynomials
\cite{wa3}-\cite{pg}, \cite{wz84}-\cite{kdec} and have also been studied in the
context of ``decorated'' spin models \cite{syozi}. As with edge inflations, it
is natural to define a uniform $\ell$-fold vertex inflation ${\cal
V}_{\ell}(G)$ as the graph obtained by inserting $\ell$ degree-2 vertices on
each edge of $G$. For the special case where $G$ is a section of a regular
lattice, it is also natural to consider edge or homeomorphic inflation of edges
forming a subset of lattice vectors. For example, for a $d$-dimensional
Euclidean lattice ${\mathbb E}^d$, one can consider the case in which the edge
or vertex inflation is performed on the edges along one or more lattice vectors
$\hat e_j$, where $j$ takes values in a subset of $\{1,...,d\}$.

\section{Relations for Uniform Edge and Vertex Inflations}

\subsection{Edge Inflation}

The effect of a uniform edge inflation on the Potts model partition function
can be determined from an analysis of the Potts Hamiltonian.  As we edge-expand
$G$ to ${\cal E}_\ell(G)$, i.e., replace each edge by $\ell$ edges joining the
same vertices, we induce the change in the Hamiltonian
\beq
{\cal H} = -J\sum_{e_{ij}} \delta_{\sigma_i \sigma_j}  \to 
           -\ell J \sum_{e_{ij}} \delta_{\sigma_i \sigma_j} \ , 
\label{hedgeinflation}
\eeq
i.e., $J \to \ell J$, and hence
\beq
y \to y_{e,\ell} = y^\ell \ ,  
\label{yedgeinflation}
\eeq
where the subscripts $e,\ell$ refer to the $\ell$-fold \underline{e}dge
inflation.  Equivalently,
\beq
v \to v_{e,\ell} = (v+1)^\ell - 1 \ , 
\label{vedgeinflation}
\eeq
where $v_{e,\ell} \equiv y_{e,\ell}-1$. This proves the following relation
connecting $Z$ on $G$ with $Z$ on ${\cal E}_\ell(G)$:
\beq
Z({\cal E}_\ell(G),q,v) = Z(G,q,v_{e,\ell}) \ . 
\label{zedgeinflation}
\eeq

We next determine the effect of this edge inflation on the Tutte
polynomial. Given the transformation (\ref{yedgeinflation}) and the fact that
$q$ does not change, so that 
\beq
q = (x-1)(y-1) = (x_{e,\ell}-1)(y_{e,\ell}-1) \ , 
\label{qinvedge}
\eeq
we have
\beq
x_{e,\ell} = 1+ \frac{(x-1)(y-1)}{y^\ell-1} = 
           1+\bigg ( \frac{x-1}{\sum_{j=0}^{\ell-1} y^j} \bigg ) \ . 
\label{xedgeinflation}
\eeq
Combining these transformations of variables with Eq. (\ref{ztrel}), we derive
the relation connecting the Tutte polynomials of $G$ and of ${\cal E}_\ell(G)$,
namely 
\beq
T({\cal E}_\ell(G),x,y) = T(G,x_{e,\ell},y_{e,\ell}) \ , 
\label{tedgeinflation}
\eeq
where $x_{e,\ell}$ and $y_{e,\ell}$ were given in Eqs. (\ref{xedgeinflation})
and (\ref{yedgeinflation}).  Note that the effect of the edge inflation on the
Potts partition function is simpler than the effect on the Tutte polynomial,
since in the former case, only one of its variables is modified, namely, $v \to
v_{e,\ell}$, whereas in the latter case, both of its variables are modified, as
$x \to x_{e,\ell}$ and $y \to y_{e,\ell}$.

Although our main focus is on the Potts model with spin-spin exchange constants
$J$ that are independent of the edges $e_{ij}$, we note parenthetically that
one can consider a generalization in which the $J_{ij}$ are different for each
edge of $G$, $e_{ij}$.  From the original 1944 Onsager solution of the
two-dimensional Ising model \cite{onsager44}, a large number of studies of spin
models have dealt with the general case with different spin-spin exchange
constants for different lattice directions. Studies of spin models with
spin-spin exchange constants $J_{ij}$ that can be different (in magnitude and
sign) for each edge, $e_{ij}$ were motivated by early work on spin glasses
\cite{spin_glasses}. In this case, where $J_{ij}$'s depend on $e_{ij}$, the
transformation of the Hamiltonian becomes
\beq
{\cal H} = -\sum_{e_{ij}} J_{ij} \delta_{\sigma_i \sigma_j}  \to 
           -\ell \sum_{e_{ij}} J_{ij} \delta_{\sigma_i \sigma_j} \ . 
\label{hijedgeinflation}
\eeq
Hence, defining $K_{ij} \equiv \beta J_{ij}$ and $v_{ij} \equiv e^{K_{ij}}-1$,
we have
\beq
v_{ij} \to v_{ij,e,\ell}=(v_{ij}+1)^\ell-1 \ . 
\label{vijedgeinflation}
\eeq
Denoting $\{v\}$ as the set of $v_{ij}$'s, we then have 
\beq
Z({\cal E}_\ell(G),q,\{v\}) = Z(G,q,\{v_{e,\ell}\}) \ . 
\label{zijedgeinflation}
\eeq

Since $J < 0$ for the Potts antiferromagnet, it follows that, as $T \to 0$, $K
\to -\infty$ (and $v \to -1$).  Hence, in this limit, the only contributions to
the partition function are from spin configurations in which adjacent spins
have different values.  The resultant $T=0$ Potts AFM partition function is
therefore precisely the chromatic polynomial $P(G,q)$ of the graph $G$ counting
the number of proper $q$-colorings of $G$: 
\beq
Z(G,q,-1) = P(G,q) \ . 
\label{zprel}
\eeq
This is equivalent, via Eq. (\ref{ztrel}), to the relation
\beq
P(G,q) = (-q)^{k_c(G)}(-1)^{n(G)}T(G,1-q,0) \ .
\label{ptrel}
\eeq
The minimum number of colors necessary for a proper $q$-coloring of $G$ is the
chromatic number $\chi(G)$.  For $q > \chi(G)$, $P(G,q)$ grows exponentially
with $n$, leading to a ground state degeneracy per vertex, $W(\{G\},q) > 1$,
where $W(\{G\},q) = \lim_{n \to \infty} P(G,q)^{1/n}$.  The ground state
entropy per vertex of the Potts model on $\{G\}$ is $S_0(\{G\},q) = k_B \ln
[W(\{G\},q)]$. In Refs. \cite{w}, \cite{wa3}-\cite{pg}, \cite{hs}, we applied
our exact results on chromatic polynomials to study the phenomenon of nonzero
ground state entropy per site in Potts antiferromagnets. These exact results
complement other approaches to studying $S_0$, such as rigorous bounds, series,
and Monte Carlo measurements \cite{bbound,wn}.  

It is clear from the definition of the chromatic polynomial that $P(G,q)$ does
not change if one replaces any edge of $G$ by two or more edges joining the
same vertices.  In particular, for the case of an $\ell$-fold uniform edge
inflation of $G$,
\beq
P({\cal E}_\ell(G),q) = P(G,q) \ . 
\label{pedgeinflation}
\eeq
This is also clear analytically from Eqs. (\ref{vedgeinflation}) and 
(\ref{zedgeinflation}), since the condition that $v=-1$ implies that 
$v_{e,\ell}=-1$. 

\subsection{Vertex Inflation}

 To analyze the effect of a vertex inflation of a graph $G$, it is again
convenient to start with the Potts model formulation.  For simplicity, we
assume here that $G$ does not have any multiple edges; it is straightforward to
extend our calculation to the case of multiple edges. We use the fact that in
the basic expression $Z = \sum_{\{ \sigma_i\} }e^{-\beta {\cal H}}$, one can
perform the summations over the spins that are located at these degree-2
vertices.  Let us consider an edge, $e_{ij}$, and insert a degree-2 vertex
$v_a$ (and its associated spin, $\sigma_a$) on this edge.  Then
\beqs
& & \sum_{\sigma_i,\sigma_a,\sigma_j} (1+v\delta_{\sigma_i \sigma_a}) 
                                      (1+v\delta_{\sigma_a \sigma_j}) \cr\cr
& &=\sum_{\sigma_i,\sigma_a,\sigma_j} \Big [1+v(\delta_{\sigma_i \sigma_a}+
           \delta_{\sigma_a \sigma_j})+ v^2\delta_{\sigma_i\sigma_j} \Big ] \
. \cr\cr 
& & 
\label{vvprod1}
\eeqs
Now, carrying out the summation over $\sigma_a$, we find that if $\sigma_i =
\sigma_j$, then the result is $q+2v+v^2$, while if $\sigma_i \ne \sigma_j$,
then the result is $q+2v$.  Therefore, 
\beqs
& & \sum_{\sigma_i,\sigma_a,\sigma_j} (1+v\delta_{\sigma_i \sigma_a}) 
                                   (1+v\delta_{\sigma_a \sigma_j}) \cr\cr
& &=(q+2v)\sum_{\sigma_i,\sigma_j}(1+v_{v,1} \delta_{\sigma_i\sigma_j}) \ , 
\label{vvprod2}
\eeqs
where
\beq
v_{v,1}= \frac{v^2}{q+2v} 
\label{vv1}
\eeq
and the subscript $v,1$ refers to the insertion of one additional vertex on the
edges. Performing this summation for each edge, we derive the relation
\beq
Z({\cal V}_1(G),q,v)=(q+2v)^{e(G)}Z(G,q,v_{v,1}) \ . 
\label{zvertexinflation1}
\eeq
This operation can be performed iteratively $\ell$ times, thereby giving a
relation between $Z$ on ${\cal V}_\ell(G)$ and $Z$ on $G$. For example, for 
$\ell=2$, we have 
\beqs
v_{v,2} & = & \frac{v_{v,1}^2}{q+2v_{v,1}} \cr\cr
        & = & \frac{v^4}{(q+2v)(q^2+2qv+2v^2)} \ , 
\label{vv2}
\eeqs
and so forth for higher values of $\ell$. We can thus relate $Z$ on ${\cal
V}_\ell(G)$ to $Z$ on $G$.

A basic problem in graph theory is the enumeration of discretized flows on the
edges of a (connected) $G$ that satisfy flow conservation at each vertex,
i.e. for which there are no sources or sinks.  One arbitrarily chooses a
direction for each edge of $G$ and assigns a discretized flow value to it.  The
value zero is excluded, since it is equivalent to the edge being absent from
$G$; henceforth, we take a $q$-flow to mean implicitly a nowhere-zero
$q$-flow. The flow on each edge can thus take on any of $q-1$ values modulo
$q$.  The flow or current conservation condition is that the flows into any
vertex must be equal, mod $q$, to the flows outward from this vertex.  These
are called $q$-flows on $G$, and the number of these is given by the flow
polynomial, $F(G,q)$.  This is a special case of the Tutte polynomial for $x=0$
and $y=1-q$ or equivalently, in terms of Potts model variables, $v=-q$: 
\beq
F(G,q) = (-1)^{c(G)} \, T(G,0,1-q) \ .
\label{ftrel}
\eeq
As is clear from the definition of the flow polynomial, adding or removing a
degree-2 vertex from an edge of $G$ does not change the number of allowed
$q$-flows on $G$, so 
\beq
F({\cal V}_\ell(G),q) = F(G,q) \ . 
\label{flowvertexinflation}
\eeq
As is evident from Eq. (\ref{vv1}), the condition $v=-q$ implies that
$v_{v,1}=-q$ and hence, more generally, that $v_{v,\ell}=-q$ for arbitrary
$\ell \ge 1$.  Recall that the number of cycles in $G$ is unchanged by this
homeomorphic inflation: $c({\cal V}_\ell(G))=c(G)$.

\section{Physical Effects of Edge and Vertex Inflation}

\subsection{General}

An important question concerns the effect of edge and vertex inflations on the
physical properties of the Potts model.  First, one may investigate these
effects for the $q$-state Potts model on (the thermodynamic limit of) a regular
lattice graph of dimension $d \ge 2$, where the ferromagnetic version of the
model has a finite-temperature order-disorder phase transition, and, depending
on the lattice type and the value of $q$, the Potts antiferromagnet may have a
finite-temperature order-disorder phase transition.  In particular, one can
study the effect on the phase transition ($pt$) temperature $T_{pt}$ for
lattices where the equation for this quantity is known exactly. However, aside
from the $q=2$ Ising case on two-dimensional lattices, there is no known exact
closed-form solution for the free energy of the $q$-state Potts model at
arbitrary temperature on these lattices with $d \ge 2$.  There is thus also
some interest in studying the effect of edge and vertex inflation for
infinite-length limits of quasi-one-dimensional lattice strips, where one can
obtain exact closed-form expressions for the free energy for arbitrary $q$ and
$T$. Of course, a spin model with short-ranged interactions, such as the Potts
model, does not have any finite-temperature phase transition on an
infinite-length quasi-one-dimensional lattice strip.  Nevertheless, one
interesting application of exact solutions for the free energy and
thermodynamic quantities on these strips is that one can study their dependence
(and the dependence of the associated Tutte polynomials) on graphical
properties, in particular, on the edge or vertex inflation.

In accordance with our notation above for the dimensionless inverse
temperature $K \equiv J/(k_BT)$, we define $K_{pt} \equiv J/(k_BT_{pt})$.  We
also introduce some notation that we will use below. We define the shifted
values of $T_{pt}$ due to a uniform $\ell$-fold edge inflation (symbolized by
the subscript $e,\ell$) and a uniform $\ell$-fold vertex inflation (symbolized
by the subscript $v,\ell$) by $T_{pt,e,\ell}$ and $T_{pt,v,\ell}$. We thus
denote $K_{pt,e,\ell} \equiv J/(k_B T_{pt,e,\ell})$ and $K_{pt,v,\ell} \equiv
J/(k_B T_{pt,v,\ell})$.

\subsection{Effect on $T_{pt}$ Due to Edge Inflation}

A general result is that for the Potts model on (an infinite) regular lattice
graph $\{G\}$ with dimensionality $d \ge 2$ that has a finite-temperature phase
transition (of either ferromagnetic or antiferromagnetic type, and either first
or second order) at a temperature $T_{pt}$, a uniform $\ell$-fold inflation of
all edges of $\{ G \}$ has the effect of multiplying $T_{pt}$ by the factor
$\ell$, i.e.,
\beq
T_{pt,e,\ell} = \ell T_{pt} \ , \quad  K_{pt,e,\ell} =  \frac{K_{pt}}{\ell} \
. 
\label{tpeell}
\eeq
Analytically, this follows because the uniform $\ell$-fold edge inflation of
$\{G\}$ changes $J$ to $\ell J$ and $T_{pt}$ is proportional to $J$.
Physically, it follows because replacing $J$ by $\ell J$ with $\ell \ge 2$
strengthens the spin-spin interaction and hence makes possible the onset of
long-range magnetic order in the presence of greater thermal fluctuations,
i.e., at a higher temperature. 

As an example, consider the Potts ferromagnet on the (thermodynamic limit of
the) square lattice, $\{sq\} = {\mathbb E}^2$.  Denoting the spin-spin exchange
constants in the two lattice directions $\hat e_i$, $i=1,2$, as $J_i$, with
$K_i = \beta J_i$ and $v_i=e^{K_i}-1$, we recall the well-known equation for
the phase transition temperature, namely \cite{wurev},
\beq
v_1 v_2 = q \ .
\label{criteq_sq}
\eeq
Let us initially assume $K_1=K_2 \equiv K$ and thus $v_1 = v_2 \equiv v$, so
Eq. (\ref{criteq_sq}) becomes $v^2=q$.  The dimensionless inverse phase 
transition temperature $K_{pt}$ is then given by 
\beq
K_{pt} = \ln( 1+\sqrt{q} \ ) \ . 
\label{kpt}
\eeq
Now let us carry out an $\ell$-fold edge inflation on the lattice. Denoting the
resultant inverse phase transition temperature in an obvious notation as 
$K_{pt,e,\ell}$, we have 
\beq
K_{pt,e,\ell} = \frac{1}{\ell} \ln (1+\sqrt{q} \ ) = \frac{K_{pt}}{\ell}  \ . 
\label{kptprime}
\eeq

One can also consider the effect of an edge inflation on all edges along a
subset of lattice directions.  The effect is simplest for the ferromagnetic
case, since this does not involves competing interactions or frustration.  This
edge inflation along a subset of lattice directions strengthens the net
spin-spin interaction and therefore makes possible the ordering associated with
the phase transition in the presence of greater thermal fluctuations.  Let us
denote $T_{pt,es,\ell}$ as the shifted phase transition temperature after an
$\ell$-fold edge inflation along a subset $s$ of the lattice directions, and
similarly denote $K_{pt,es,\ell} \equiv J/(k_B T_{pt,es,\ell})$. Then the
reasoning above yields the inequality $T_{pt,es,\ell} > T_{pt}$ for $\ell \ge
2$. As an example, we again consider the (infinite) square lattice $\{ sq \}$
and perform an edge inflation with $\ell=2$ for edges in one of the two lattice
directions, say $\hat e_2$, so that $K_1 = K$, $K_2 = 2K$.  The equation for
the shifted inverse phase transition temperature is given by $v^2(v+2)=q$, with
physical solution 
\beq
K_{pt,e2,2} =  \ln \bigg [ \frac{1}{3} \Big \{ A^{1/3} + 4A^{-1/3} + 1 
\Big \} \bigg
] \ , 
\label{ekptsqonedir}
\eeq
where the subscript $e2$ means inflation along edges along $\hat e_2$ and 
\beq
A = \frac{1}{2} \Big [ 27q-16+3\sqrt{3q(27q-32)} \ \Big ] \ . 
\label{aedgeonedir}
\eeq
The resultant $K_{pt,e2,2} < K_{pt}$, i.e., $T_{pt,e2,2} > T_{pt}$, 
in agreement with the general argument given above.  For
example, for $q=2$, $K_{pt} = \ln(1+\sqrt{2} \ ) \simeq 0.88137$, while 
$K_{pt,e2,2}$ is given by 
\beq
e^{K_{pt,e2,2}} = \frac{1}{3} \bigg [ (19+3\sqrt{33} \ )^{1/3} + 
4(19+3\sqrt{33} \ )^{-1/3} + 1 \bigg ] \ , 
\label{ekptprime}
\eeq
so that $K_{pt,e2,2} \simeq 0.60938$.  

\subsection{Effect on $T_{pt}$ Due to Vertex Inflation}

One can also deduce a general result for the effect of a uniform $\ell$-fold
vertex (i.e., homemorphic) inflation of the thermodynamic limit of a lattice
graph $\{G\}$ with dimensionality $d \ge 2$.  A generic feature of the phase
transition temperature $T_{pt}$ of a ferromagnetic spin model (above its lower
critical dimensionality, so that this temperature is finite) is that, other
things being equal, $T_{pt}$ increases as a function of the vertex degree,
i.e., coordination number, of the lattice.  This feature is observed in
approximate determinations of $T_{pt}$ from high-temperature and
low-temperature series expansions, Monte Carlo simulations, mean-field
approximations, and, where available, exact solutions for $T_{pt}$.  It is
understood physically as a consequence of the fact that increasing the
coordination number increases the effect of the spin-spin interactions, so that
the ordering associated with the phase transition can occur in the presence of
greater thermal fluctuations.  On an (infinite) line, with coordination number
$\kappa = 2$, a spin model with short-range interactions does not have a
finite-temperature phase transition, so the lattices of interest in this
section are lattices with $d \ge 2$, which necessarily have coordination number
$\kappa \ge 3$ (where this minimum value, $\kappa=3$, is realized for the
honeycomb lattice).  If one starts with a lattice graph $\{G\}$ with
coordination number $\kappa \ge 3$, then a uniform vertex inflation, which
consists of the addition of $\ell$ degree-2 vertices on each edge of $\{ G \}$,
reduces the effective vertex degree, $\kappa_{eff}$. 

Let us, for technical simplicity, consider the thermodynamic limit $\{G\}$ to
be reached as the $n(G) \to \infty$ limit of a regular lattice graph $G$ with
periodic boundary conditions, and uniform vertex degree $\kappa$.  Then, with
$e(G) = (\kappa/2)n(G)$, the number of vertices and edges of ${\cal V}_\ell(G)$
are
\beq
n({\cal V}_\ell(G)) = \bigg ( 1 + \frac{\kappa \ell}{2} \bigg ) n(G)
\label{nvertexinflation}
\eeq
and
\beq
e({\cal V}_\ell(G)) = (1+\ell)e(G)
\label{evertexiflation}
\eeq
so that 
\beq
\kappa_{eff}({\cal V}_\ell(G)) = \bigg [ 
\frac{1+ \ell}{1+ \frac{\kappa \ell }{2}} \bigg ] \kappa \ . 
\label{kappaeffvertexinflation}
\eeq
The fact that the vertex inflation reduces $\kappa_{eff}$ if $\kappa > 2$ is
clear analytically from this result, since 
\beq
\kappa_{eff}({\cal V}_\ell(G)) < \kappa \quad {\rm if} \ \ell \ge 1 \quad {\rm
  and} \quad \kappa > 2 \ . 
\label{kappaeffinequality}
\eeq
For example, for the square lattice, with $\kappa =4$, one has
$\kappa_{eff}({\cal V}_1(sq)) = 8/3 = 2.666..$, $\kappa_{eff}({\cal V}_2(sq)) =
12/5 = 2.4$, $\kappa_{eff}({\cal V}_3(sq)) = 16/7 \simeq 2.286$, and so forth,
with an approach to 2 from above as $\ell \to \infty$. Indeed, in general, 
for an arbitrary regular lattice graph $G$ with coordination number $\kappa$, 
\beq
\lim_{\ell \to \infty} \kappa_{eff} = 2
\label{kappalimit}
\eeq
For $\ell >> 1$, one has the Taylor series expansion
\beq
\kappa_{eff} = 2 \Bigg [ 1 + \bigg ( 1- \frac{2}{\kappa} \bigg ) \bigg \{
\frac{1}{\ell} + \frac{2}{\kappa \ell^2} + O \Big ( \frac{1}{\ell^3} \Big ) 
\bigg \} \Bigg ]
\label{kappaefftaylor}
\eeq
Hence, for a Potts ferromagnet on a lattice graph $\{ G \}$ with $d \ge 2$, a
uniform $\ell$-fold vertex inflation of $\{ G\}$ with $\ell \ge 1$ leads to a
decrease in the phase transition temperature.  The same conclusion holds if one
performs a vertex inflation on all edges along a subset of the lattice
directions.

We illustrate the effect of vertex inflation for the $q$-state Potts
ferromagnet on the square lattice.  Let us perform a uniform vertex inflation
with $\ell=1$ on all edges of the lattice, i.e., add one degree-2 vertex to
each edge of this lattice.  Then, combining Eqs. (\ref{vedgeinflation}) and
(\ref{criteq_sq}), we find that the equation for the phase transition
temperature is $v_{v,1} = \sqrt{q}$, where $v_{v,1}$ was given in
Eq. (\ref{vv1}). The solution for the inverse phase transition temperature
$K_{pt,v,1}$ is
\beq
K_{pt,v,1} = \ln \bigg [ 1 + \sqrt{q} \, 
\Big \{ 1+\sqrt{1+\sqrt{q}} \ \Big \} \bigg ] \ .  
\label{kptprime_homeo}
\eeq
Comparing this with the inverse critical temperature for the original $\{ G\}$,
$K_{pt} = \ln(1 + \sqrt{q} \ )$, we see that $K_{pt,v,1} > K_{pt}$, in
agreement with our general argument above.

The situation is more complicated with the Potts antiferromagnet; we will show
that vertex inflation can either lower or raise a phase transition (critical)
temperature, depending on the value of $q$ and the lattice type.  First,
consider the $q=2$ (Ising) Potts antiferromagnet on a bipartite lattice
$\{G\}$. The bipartite property of $\{G\}$ means that it can be expressed as
the union of an even and an odd sublattice, $\{G\} = \{ G_1 \} \cup \{G_2 \}$,
with the property that each vertex in $G_1$ has, as its only adjacent vertices,
members of the vertex set of $\{G_2 \}$ and vice versa.  There is a well-known
isomorphism that maps the Ising antiferromagnet on $\{G\}$ to an Ising
ferromagnet on $\{G\}$, namely the simultaneous replacement $J \to -J$ and
$\sigma_{v_1} \to -\sigma_{v_1}$, with $\sigma_{v_2}$ unchanged, where here
$v_1$ and $v_2$ denote vertices in $G_1$ and $G_2$, respectively.  Thus, in
this case, the effect of an $\ell$-fold vertex inflation on all edges is the
same for the antiferromagnetic case as for the ferromagnetic case discussed
above, namely that it reduces $T_{pt}$.  However, we next prove that vertex
inflation can also have the opposite effect, of increasing a critical
temperature.  For this purpose, let us consider the $q=2$ Potts (Ising)
antiferromagnet on the infinite triangular lattice $\{tri\}$ (with equal
negative spin-spin exchange constants in each of the three lattice directions,
$J_1 = J_2 = J_3 \equiv J < 0$).  This antiferromagnet is frustrated, and is
only critical at $T=0$ \cite{stephenson}.  Now let us perform a uniform
$\ell$-fold vertex inflation on all edges, with $\ell=2k+1$ odd, thereby
obtaining $\{ {\cal V}_{2k+1}(tri) \}$. This lattice, $\{ {\cal V}_{2k+1}(tri)
\}$, is bipartite, in contrast with the triangular lattice itself.  Because of
this, the Ising antiferromagnet is not frustrated on $\{ {\cal V}_{2k+1}(tri)
\}$, and therefore one can apply a standard Peierls-type argument to infer that
it has a finite-temperature symmetry-breaking phase transition. Indeed, because
$\{ {\cal V}_{2k+1}(tri) \}$ is bipartite and because of the isomorphism
mentioned above, the Ising ferromagnet and antiferromagnet can be mapped to
each other, and have their respective phase transitions at the same $T_{pt}$.
Thus, as this example shows, in contrast with the situation for the Potts
ferromagnet, vertex inflation for the Potts antiferromagnet may actually raise
a critical or phase transition temperature rather than lowering it, depending
on $q$ and the lattice type.

\subsection{Invariance of Universality Class Under Edge Inflations} 

We recall that on two-dimensional lattices the (zero-field) $q$-state Potts
ferromagnet with $q \le 4$ has a second-order phase transition with an
associated $q$-dependent universality class and corresponding thermal and
magnetic critical exponents that are independent of the lattice type.  It is of
interest to study whether vertex or edge inflation changes the universality
class of this phase transition.  A general result of renormalization-group
analyses of second-order phase transitions is that the universality class of a
second-order phase transition (in a model that is free of complications such as
competing interactions, frustration, and/or quenched disorder) depends on the
lattice dimensionality, $d$ and the symmetry group of the Hamiltonian (e.g.,
\cite{rg}).  For the Potts model, the symmetry group of the Hamiltonian is the
symmetric (permutation) group on $q$ indices, $S_q$.  Edge 
inflations of the lattice have no effect on the dimensionality or symmetry
group of the Hamiltonian; hence, for the range $q \le 4$ where the
two-dimensional Potts ferromagnet has a second-order phase transition, they do
not change the universality class of this transition.

\subsection{Effects of Vertex Inflation on Universality Class} 

We again consider the interval $q \le 4$ where the two-dimensional Potts
ferromagnet has a second-order phase transition.  As with edge inflation, we
note that vertex inflation has no effect on the lattice dimensionality or
symmetry group of the Hamiltonian, and hence, by the same argument as before,
it does not change the universality class of the phase transition.

As before, the situation is more complicated for the Potts antiferromagnet,
because, in contrast to the ferromagnet, its properties depend sensitively on
the type of lattice. To show this, it is convenient to continue with the same
illustrative example that we used above, namely the $q=2$ Ising
antiferromagnet.  We will give two examples which exhibit opposite behaviors;
in the first, the vertex inflation lowers the phase transition temperature and
makes no change in the universality class.  In the second, the vertex inflation
raises the critical temperature and does change the universality class.  The
first example uses the Ising antiferromagnet on a bipartite lattice graph
$\{G\}$ of dimensionality $d \ge 2$, where it has a phase transition at a
temperature $T_{pt}$ (with antiferromagnetic long-range order for $T <
T_{pt}$).  Now let us perform a uniform $\ell$-fold vertex inflation on all of
the edges of $\{G\}$, thereby obtaining $\{ {\cal V}_\ell(G) \}$. By an
argument similar to the one given above, this lowers $T_{pt}$.  Since this
vertex inflation does not change either the lattice dimensionality or the
symmetry group $S_2 \approx {\mathbb Z}_2$, it leaves the universality class
unchanged. 

To show that the opposite can also happen, let us consider the (isotropic)
Ising antiferromagnet on the triangular lattice.  As mentioned before, because
of the frustration, this model has no finite-temperature transition, but is
critical at $T=0$; the spin-spin correlation function decays asymptotically
like $\langle \sigma_0 \sigma_{\vec r} \rangle \propto r^{-1/2} \cos(2 \pi
r/3)$ for large $r = |\vec r|$ \cite{stephenson}.  Normally, at a second-order
phase transition of a spin model on a $d$-dimensional lattice in which the
(connected) spin-spin correlation function decays asymptotically like $\langle
\sigma_0 \sigma_{\vec r} \rangle \propto r^{-(d-2+\eta})$, one assigns the
critical exponent $\eta$ to this transition.  Because of the oscillatory nature
of the asymptotic decay of the $T=0$ Ising antiferromagnet on the triangular
lattice, this case is more complex, but the decay of the envelope curve is
described by $\eta=1/2$.  Now, just as we did before, let us perform a uniform
$\ell$-fold vertex inflation on all edges, with $\ell=2k+1$ odd, thereby
obtaining the bipartite lattice $\{ {\cal V}_{2k+1}(tri) \}$. As discussed
above, because $\{ {\cal V}_{2k+1}(tri) \}$ is bipartite, the Ising
antiferromagnet is not frustrated on it, and, indeed, can be mapped to the
Ising ferromagnet by the mapping given in the previous subsection. Owing to
this, the Ising ferromagnet and antiferromagnet on this lattice have the same
phase transition temperatures, and, furthermore, the Ising antiferromagnet is
automatically in the same universality class as the Ising ferromagnet, with
$\eta=1/4$ \cite{rg,eta_ising}. Thus, in this case, the vertex inflation does
change both the value of the critical temperature and the universality class of
the phase transition.

\section{Longitudinal Homeomorphic Inflations of Free Ladder Graph}

In the previous sections we have derived general relations that connect
$Z(G,q,v)$ and $Z(\tilde G, q,v)$, where $\tilde G$ is obtained from $G$ by
uniform edge or vertex inflations.  By Eq. (\ref{ztrel}), these enable one to
calculate the equivalent Tutte polynomial, $T(\tilde G,x,y)$.  We have also
discussed the case where edge or vertex inflations are performed on all edges
along a subset of lattice directions of a lattice graph.  In the rest of this
paper we explore the latter type of vertex inflation further.  We present
exact calculations of Potts/Tutte polynomials for a class of vertex (i.e.,
homeomorphic) inflations of a subset of the edges of ladder graphs. Our results
generalize our calculations of the chromatic polynomials for these graphs with
S.-H. Tsai in Refs. \cite{pg} and \cite{wa3} (see also \cite{hs,wa2}).

We consider the free strip of the ladder graph comprised of $m$ squares, which
we denote as $S_m$.  Now we perform an $\ell$-fold vertex inflation on all of
the longitudinal edges, with $\ell=k-2$ and $k \ge 3$, i.e., we insert $k-2$
degree-2 vertices on each of these edges.  The resultant strip graph is denoted
$S_{k,m}$.  The transverse edges are not affected by this operation.  The
original ladder graph itself is $S_{2,m}$.  The numbers of vertices and edges
on $S_{k,m}$ are
\beq
n(S_{k,m})=2(k-1)m+2
\label{nskm}
\eeq
and
\beq
e(S_{k,m})=(2k-1)m+1 \ . 
\label{eskm}
\eeq

Using a systematic iterative application of the deletion-contraction property,
we calculate the Tutte polynomial, $T(S_{k,m},x,y)$. One way to express this is
in terms of a generating function.  We use a generating function
\beq
\Gamma(S_k,x,y;z) = \sum_{m=0}^\infty T(S_{k,m+1},x,y)z^m 
\label{gamdef}
\eeq
of the form 
\beq
\Gamma(S_k,x,y;z) =\frac{a_{0}+a_{1}z}{1 + b_{1}z + b_{2} z^2} \ . 
\label{gammaform}
\eeq
The denominator can be written as 
\beq
1 + b_{1}z + b_{2} z^2 = (1-\lambda_{_{k,0,1}}z) (1-\lambda_{_{k,0,2}}z) \ .
\label{gden}
\eeq
Recall that for the circuit graph with $n$ vertices, $C_n$, 
\beq
T(C_n,x,y) = \frac{x^n + c^{(1)}}{x-1} = y + \sum_{j=1}^{n-1} x^j \ , 
\label{tcn}
\eeq
where $c^{(1)}=q-1=xy-x-y$. We calculate 
\beq
a_{0}=T(C_{2k},x,y) \ , 
\label{a0}
\eeq
\medskip
\beq
a_{1}=-yx^{2k-1} \ , 
\label{a1}
\eeq
\beq
b_{0} = -[ 1+T(C_{2k-1},x,y) ] \ , 
\label{b0}
\eeq
and
\beq
b_{1} = x^{2k-2}y \ . 
\label{b1}
\eeq
Thus, 
\beq
\lambda_{k,0,j} = \frac{1}{2}\Big [-b_{0} \pm \sqrt{b_{0}^2-4b_{1}} \ \Big ] 
\ . 
\label{lamk0j}
\eeq
By a generalization of Eq. (2.15) in \cite{hs} from chromatic polynomials to
the full Tutte polynomial, it follows that 
\begin{widetext}
\beq
T(S_{k,m},x,y) = \frac{(a_{0}\lambda_{k,0,1}+a_{1})}
{(\lambda_{k,0,1}-\lambda_{k,0,2})} \, (\lambda_{k,0,1})^{m-1}
+ \frac{(a_{0}\lambda_{k,0,2}+a_{1})}
{(\lambda_{k,0,2}-\lambda_{k,0,1})} \, (\lambda_{k,0,2})^{m-1} \ .
\label{tskm}
\eeq
\end{widetext}
Note that $T(S_{k,m},x,y)$ is symmetric under the interchange $\lambda_{k,0,1}
\leftrightarrow \lambda_{k,0,2}$.  It is straightforward, using
Eq. (\ref{ztrel}), to re-express these results in terms of the Potts model
partition function $Z(S_{k,m},q,v)$; for brevity, we omit the explicit results.

\section{Longitudinal Homeomorphic Inflations of Cyclic and M\"obius Ladder 
Graphs}

In this section we present an exact calculation of the Tutte polynomial for
longitudinal homeomorphic inflations of cyclic and M\"obius ladder graphs. We
start with a cyclic ladder strip of length $m$ squares and add $k-2$ degree-2
vertices, with $k \ge 3$, to each longitudinal edge.  This yields the strip
graph with longitudinal homeomorphic inflation that we denote $L_{k,m}$. The
corresponding M\"obius strip $M_{k,m}$ is obtained by cutting the cyclic graph
at any transverse edge and reattaching the ends after a vertical twist.
The graphs $L_{k,m}$ and $M_{k,m}$ each have the number of vertices 
\beq
n(L_{k,m}) = n(M_{k,m}) = 2(k-1)m 
\label{nlkm}
\eeq
and the number of edges 
\beq
e(L_{k,m}) = e(M_{k,m}) = (2k-1)m \ . 
\label{elkm}
\eeq
For a given $m$ and for $k=2$ these are the original cyclic and M\"obius ladder
strip graphs; the case $k=3$ is the first homeomorphic inflation of these
respective graphs, and so forth for higher values of $k$. 

For the cyclic strip graph $L_{k,m}$ we calculate
\beq
T(L_{k,m},x,y) = \frac{1}{x-1}\sum_{d=0}^2 c^{(d)} \sum_{j=1}^{n_T(2,d)} 
(\lambda_{k,d,j})^m
\label{tlad}
\eeq
where $c^{(0)}=1$, $c^{(1)}=q-1$, $c^{(2)}=q^2-3q+1$, 
$n_T(2,0)=2$, $n_T(2,1)=3$, and $n_T(2,2)=1$.  The 
$\lambda_{k,0,j}$ for $j=1,2$ were given above in Eq. (\ref{lamk0j}). For the
others, we find 
\beq
\lambda_{k,1,1} = x^{k-1} \ , 
\label{lamhomd1j1}
\eeq
\beq
\lambda_{k,1,j} = \frac{1}{2}( u_k \pm \sqrt{r_k} \  ) \ , \quad j=2, \ 3 \ , 
\label{lamhomd1j23}
\eeq
where 
\beqs
u_k & = & 2\Big ( \sum_{j=0}^{k-2}x^j \Big ) + x^{k-1} + y \cr\cr
    & = &  \frac{x^k+x^{k-1}-2}{x-1} + y \ , 
\label{uk}
\eeqs
and
\beq
r_k = u_k^2-4x^{k-1}y \ , 
\label{rk}
\eeq
and finally, 
\beq
\lambda_{k,2,1}=1 \ . 
\label{lamhomd2}
\eeq
Although our result (\ref{tlad}) (and (\ref{tmob}) below) are formally rational
functions in $x$, one easily verifies that the prefactor $1/(x-1)$ divides the
expression to its right, so that $T(L_{k,m},x,y)$ and $T(M_{k,m},x,y)$ are
polynomials in $x$ as well as $y$, as guaranteed by Eq. (\ref{t}).  It is again
straightforward, using Eq. (\ref{ztrel}), to re-express these results in terms
of the Potts model partition function $Z(S_{k,m},q,v)$.

For the M\"obius strip $M_{k,m}$, we find that the $\lambda$'s are the same as
for the cyclic strip $L_{k,m}$, and the pattern of changes in the coefficients
is the same as was shown in \cite{pm,a,cf} for lattice strips 
without homeomorphic expansion, so that
\beqs
& & T(M_{k,m},x,y) = \frac{1}{x-1}\Bigg [\sum_{j=1}^2 (\lambda_{k,0,j})^m 
\cr\cr
& & 
+ c^{(1)}\Big ( - (\lambda_{k,1,1})^m + \sum_{j=2}^3 (\lambda_{k,1,j})^m 
\Big ) - 1 \Bigg ] \ . \cr\cr
& & 
\label{tmob}
\eeqs
From these general expressions, one can specialize to the chromatic polynomials
and the flow polynomials.  Via Eq. (\ref{ptrel}), one readily checks that the
results for the chromatic polynomials agree with those that we calculated
before with S.-H. Tsai in Ref. \cite{pg}. The flow polynomials are unaffected
by homeomorphic expansion, and hence coincide with those that we calculated
with S.-C. Chang in Ref. \cite{fp}.

\section{Valuations of Tutte Polynomials for Homeomorphic Expansions of 
Cyclic and M\"obius Ladder Graphs}

Special valuations of the Tutte polynomial yield several quantities of
graph-theoretic interest.  In this section we calculate these for the
longitudinal homeomorphic expansions of cyclic and M\"obius ladder graphs.  We
first recall some definitions.  A tree graph is a connected graph with no
circuits (cycles).  A spanning tree of a graph $G$ is a spanning subgraph of
$G$ that is also a tree.  A spanning forest of a graph $G$ is a spanning
subgraph of $G$ that may consist of more than one connected component but
contains no circuits. The special valuations of interest here are (i)
$T(G,1,1)=N_{ST}(G)$, the number of spanning trees ($ST$) of $G$; (ii)
$T(G,2,1)=N_{SF}(G)$ the number of spanning forests ($SF$) of $G$; (iii)
$T(G,1,2)=N_{CSSG}(G)$, the number of connected spanning subgraphs ($CSSG$) of
$G$; and (iv) $T(G,2,2)=N_{SSG}(G)=2^{e(G)}$, the number of spanning subgraphs
($SSG$) of $G$.  The last of these quantities is determined directly from
Eq. (\ref{elkm}) as
\beq
N_{SSG}(G_{k,m}) = 2^{(2k-1)m} \ , 
\label{nssg}
\eeq
where we introduce the notation $G_{k,m}$ to stand for either $L_{k,m}$ or
$M_{k,m}$.  

We calculate
\beq
\lambda_{k,0,j}{}\Big |_{x=2 \atop y=1}=2^{k-1}\Big [ 2^{k-1} \pm 
\Big (2^{2(k-1)}-1 \Big )^{1/2} \, \Big ] \ , 
\label{lamhomd0j12_x2y1}
\eeq
where the $\pm$ sign applies for $j=1,2$, respectively, 
\beq
\lambda_{k,1,1}{}\Big |_{x=2 \atop y=1}=2^{k-1} \ , 
\label{lamhomd1j1_x2y1}
\eeq
and
\beqs
& & \lambda_{k,1,j}{}\Big |_{x=2 \atop y=1} = \cr\cr
& & \frac{1}{2}\bigg [ 3 \cdot 2^{k-1}-1 \pm 
\Big [ (3 \cdot 2^{k-1}-1)^2-2^{k+1} \Big ]^{1/2} \bigg ] \ , \cr\cr
& &
\label{lamhomd1j23_x2y1}
\eeqs
where the $\pm$ sign applies for $j=2,3$, respectively. 
Next, define the notation 
\beq
\eta_G \equiv \eta_{G_{k,m}}=\begin{cases} +1 & \text{if $G_{k,m}=L_{k,m}$}\cr
                                           -1 & \text{if $G_{k,m}=M_{k,m}$} 
\end{cases}
\label{eta}
\eeq
In terms of these quantities, we have, for $G_{k,m}=L_{k,m}$ or $M_{k,m}$, 
\beqs
& & N_{SF}(G_{k,m})=(\lambda_{k,0,1})^m+(\lambda_{k,0,2})^m
\cr\cr
& & + \eta_G \Big [ 1-(\lambda_{k,1,1})^m \Big ] - 
             \Big [ (\lambda_{k,1,2})^m+(\lambda_{k,1,3})^m \Big ] \cr\cr
& & 
\label{tgkmx2y1}
\eeqs
where in Eq. (\ref{tgkmx2y1}) the $\lambda_{k,d,j}$'s are evaluated at $x=2$,
$y=1$ as in Eqs.  (\ref{lamhomd0j12_x2y1})-(\ref{lamhomd1j23_x2y1}).  For
$k=2$, i.e. the original cyclic and M\"obius strips, one checks that
eq. (\ref{tgkmx2y1}) reduces to eq. (D.14) of our previous work \cite{a}.  As
an example, for the first homeomorphic expansion, $k=3$, eq. (\ref{tgkmx2y1})
yields
\beqs
& & N_{SF}(G_{3,m})= \cr\cr
& & [4(4+\sqrt{15}\ )]^m + [4(4-\sqrt{15}\ )]^m + \eta_G(1-2^{2m}) \cr\cr
& &  - \bigg [ \bigg ( \frac{11+\sqrt{105}}{2} \ \bigg )^m 
+ \bigg ( \frac{11-\sqrt{105}}{2} \ \bigg )^m \bigg ] \ . 
\label{tgk3mx2y1}
\eeqs

For valuations of $T(G_{k,m},x,y)$ with $x=1$ and $G_{k,m}=L_{k,m}$ or
$M_{k,m}$, a useful equality is
\begin{widetext}
\beq
\lambda_{k,0,j}{}\Big |_{x=1} = \lambda_{k,1,j+1}{}\Big |_{x=1} 
=   
\frac{1}{2}\bigg [2k-1+y \pm \Big [ (2k-1+y)^2-4y \Big ]^{1/2} \, \bigg ] \ , 
\label{lamhomd0j_x1}
\eeq
where the $\pm$ sign applies for $j=1,2$, respectively.  For the number of
spanning trees on these families on $G_{k,m}=L_{k,m}$ or $M_{k,m}$, we find
\beq
N_{ST}(G_{k,m})= m(k-1)\bigg [ -\eta_G  
+ \frac{1}{2}\Big \{ \Big (k+\sqrt{k^2-1} \ \Big )^m +  
                         \Big (k-\sqrt{k^2-1} \ \Big )^m \Big \} \bigg ] \ . 
\label{tgkmx1y1} 
\eeq
\end{widetext} 
For $k=2$, this reduces to Eq. (D.13) of \cite{a}. 

Next, we introduce the shorthand notation
\beq
\lambda_{\pm} = \frac{1}{2}\Big ( 2k+1 \pm \sqrt{4k^2+4k-7} \ \Big )
\label{lampm}
\eeq
and
\beq
a_{\pm} = \Big ( \frac{k-1}{2} \Big )
\bigg [ k \pm \frac{(2k^2+k-4)}{\sqrt{4k^2+4k-7}} \ \bigg ] \ . 
\label{lamprimepm}
\eeq
In terms of these quantities, we find, for the number of connected spanning 
subgraphs, 
\beqs
& & N_{CSSG}(L_{k,m})=-2 + (\lambda_+)^m+(\lambda_-)^m \cr\cr
& + & m\bigg [ 1 - k + (\lambda_+)^{m-1}a_+ 
                     + (\lambda_-)^{m-1}a_-  \bigg ] 
\label{tgkmcycx1y2}
\eeqs
and 
\beq
N_{CSSG}(M_{k,m}) = N_{CSSG}(L_{k,m})+1+2(k-1)m \ . 
\label{tgkmmobx1y2}
\eeq
For $k=2$, this reduces to Eq. (D.15) of \cite{a}. 

These graphical quantities grow exponentially as a function of the strip length
$m$ and hence also $n$. For each quantity $N_{\cal G}$ (i.e., $N_{ST}$,
$N_{SF}$, $N_{CSSG}$, and $N_{SSG}$), one can thus define a growth constant
\beq
w_{N_{{\cal G}(G)}} \equiv \lim_{m \to \infty} [N_{{\cal G}(G)}]^{1/n} \ . 
\label{wdef}
\eeq
In Ref. \cite{a} we used an equivalent quantity to describe the exponential
growth, viz., 
\beq
z_{N_{{\cal G}(G)}} \equiv \ln [w_{N_{{\cal G}(G)}}] \ . 
\label{zw}
\eeq

We find that for a given set of subgraphs ${\cal G}(G)$, these growth constants
are the same for the $m \to \infty$ limits of the $S_{k,m}$, $L_{k,m}$, and
$M_{k,m}$ families of strip graphs, i.e.,
\beq
w_{N_{\cal G}(\{ S_k \})} = w_{N_{\cal G}(\{L_k\})} = 
w_{N_{\cal G}(\{ M_k)\}} \ , 
\label{wequality}
\eeq
so we will just label them by $\{L_k \}$. 
For each type of subgraph ${\cal G}$, the growth constant is 
determined by the dominant $\lambda_{k,d,j}$,
which is $\lambda_{k,0,1}$, evaluated at the respective values of $(x,y)$.
We find
\beq
w_{_{SSG(\{L_k\})}} = 2^{\frac{2k-1}{2(k-1)}} \ , 
\label{wssglk}
\eeq
\beq
w_{_{SF(\{L_k\})}} = 
2 \Big [1+\Big (1-2^{-2(k-1)} \Big )^{1/2} \, \Big ]^{\frac{1}{2(k-1)}} \ , 
\label{wsflk}
\eeq
\beq
w_{_{CSSG(\{L_k\})}}=
\bigg [\frac{2k+1+(4k^2+4k-7)^{1/2}}{2} \, \bigg ]^{\frac{1}{2(k-1)}} \ , 
\label{wcssglk}
\eeq
and
\beq
w_{_{ST(\{L_k\})}} = \Big ( k + \sqrt{k^2-1} \, \Big )^{\frac{1}{2(k-1)}} \ .
\label{wstlk}
\eeq
For $k=2$, i.e., the original ladder strip without homeomorphic expansion,
Eqs. (\ref{wssglk}), (\ref{wsflk}), (\ref{wcssglk}), and (\ref{wstlk}),
together with (\ref{zw}), reduce to Eqs. (D.24), (D.22), (D.23), and (D.21) of
our previous paper, Ref. \cite{a}.  Some numerical values of these growth
constants obtained from the analytic results given above are displayed in Table
\ref{wvalues}.  

\begin{table}
\caption{\footnotesize{Values of $w_{_{SSG}}$, $w_{_{SF}}$, $w_{_{CSSG}}$, and
$w_{_{ST}}$ for the $m \to \infty$ limit of the $G_{k,m}$ family of graphs,
where $G_{k,m}$ denotes $S_{k,m}$, $L_{k,m}$, or $M_{k,m}$.}}
\begin{center}
\begin{tabular}{|c|c|c|c|c|}
\hline\hline $k$ & $w_{_{SSG}}$ & $w_{_{SF}}$ & $w_{_{CSSG}}$ & $w_{_{ST}}$ 
\\ \hline\hline
2  &  2.8284271  &  2.7320508  &  2.1357792  &  1.9318517  \\ \hline
3  &  2.3784142  &  2.3689169  &  1.6089554  &  1.5537740  \\ \hline
4  &  2.2449241  &  2.2434544  &  1.4360948  &  1.4104463  \\ \hline
5  &  2.1810155  &  2.1807485  &  1.3466467  &  1.3318300  \\ \hline
6  &  2.1435469  &  2.1434946  &  1.2908358  &  1.2811894  \\ \hline
7  &  2.1189262  &  2.1189154  &  1.2522226  &  1.2454451  \\ \hline
8  &  2.1015133  &  2.1015110  &  1.2236924  &  1.2186716  \\ \hline
9  &  2.0885476  &  2.0885471  &  1.2016292  &  1.1977615  \\ \hline
10 &  2.0785185  &  2.0785183  &  1.1839857  &  1.1809157  \\ \hline
$\infty$ &  2     &  2           &   1         &   1         \\ \hline\hline
\end{tabular}
\end{center}
\label{wvalues}
\end{table}

\section{Tutte Polynomials of Hammock Graphs}

\subsection{General Calculation} 

A hammock graph $H_{k,r}$ is defined as follows: start with two vertices
connected by $r$ edges (where $r$ denotes ``rope'' or, more abstractly,
``route'').  Now add $k-2$ degree-2 vertices to each of these edges, with $k
\ge 3$, so that on any rope there are a total of $k$ vertices, including the
two end-vertices.  Thus, $H_{k,r}$ with $k \ge 3$ is a uniform $\ell=k-2$ fold
vertex (homeomorphic) inflation of the original graph, $H_{2,r}$ with $r$ edges
connecting the two end vertices. The number of vertices, edges, and (linearly
independent) cycles in this graph are
\beq
n(H_{k,r})=2+(k-2)r \ ,
\label{nhkr}
\eeq
\beq
e(H_{k,r})=(k-1)r \ ,
\label{ehkr}
\eeq
and
\beq
c(H_{k,r})=r-1 \ . 
\label{chkr}
\eeq
(satisfying the general relation $c(G)=e(G)+k_c(G)-n(G)$).  Hence, 
\beq
\kappa_{eff}(H_{k,r}) = \frac{2r(k-1)}{2+(k-2)r} \ .
\label{kappaeff_hkr}
\eeq
An interesting feature of the $H_{k,r}$ graphs is that if $r \to \infty$ for
fixed $k$, the combination of tne two end vertices, each with $\kappa=r$ and
the $(k-2)r$ interior vertices on the ``ropes'', each with $\kappa=2$, yields
the result
\beq
\lim_{r \to \infty}\kappa_{eff}(H_{k,r}) = 2 \bigg (\frac{k-1}{k-2}\bigg ) \ .
\label{kappaeff_hkr_larger}
\eeq
Although $H_{k,r}$ is much simpler than the complex networks encountered in
biological and social contexts, the property that for large $r$ it has vertices
of quite different degrees is also observed in complex networks
\cite{complexnets}.  The girth $g(G)$ of a graph $G$ is the number of edges in
a minimum-distance circuit in $G$.  For $r \ge 2$, the girth of $H_{k,r}$ is
\beq
g(H_{k,r}) = 2k-2 \ . 
\label{girth_hkr}
\eeq

Before presenting our general results for $T(H_{k,r},x,y)$, we note two special
cases. First, for $r=2$, the graph $H_{k,2}$ is just the circuit graph
with $2k-2$ vertices: 
\beq
H_{k,2} = C_{2k-2} \ . 
\label{hkr2}
\eeq
Hence, 
\beq
T(H_{k,2},x,y) = T(C_{2k-2},x,y) \ ,
\label{thkr2}
\eeq
where $T(C_n,x,y)$ was given in Eq. (\ref{tcn}).  The second special case is
for $k=2$. If $G$ is a planar graph, we denote its planar dual as $G^*$.  From
Eq. (\ref{t}), it follows that 
\beq
T(G,x,y) = T(G^*,y,x) \ . 
\label{tdual}
\eeq
We can apply this result here, since $H_{k,r}$ is a planar graph. Now 
$H_{2,r}$ is the graph consisting of two vertices connected by $r$
edges.  This is the planar dual to the circuit graph; 
\beq
H_{2,r} = (C_r)^* \ . 
\label{hk2rdualcr}
\eeq
From Eqs. (\ref{tdual}), (\ref{hk2rdualcr}), and (\ref{tcn}), it follows that 
\beq
T(H_{2,r},x,y) = T(C_r,y,x) = \frac{y^r +xy-x-y}{y-1} \ . 
\label{thk2r}
\eeq

Proceeding to the general $H_{k,r}$ graph, we define 
\beq
\lambda_{H,1}=\sum_{j=0}^{k-2} x^j = \frac{x^{k-1}-1}{x-1} 
\label{lamh1}
\eeq
and
\beq
\lambda_{H,2}=T(C_{k-1},x,y) \ . 
\label{lamh2}
\eeq
Using an iterative application of the deletion-contraction relation, we
calculate the Tutte polynomial of $H_{k,r}$ to be
\begin{widetext}
\beq
T(H_{k,r},x,y) = (\lambda_{H,1})^{r-2} \, T(C_{2k-2},x,y) 
+ \bigg [ \frac{(\lambda_{H,1})^{r-2}-(\lambda_{H,2})^{r-2}}
{\lambda_{H,1}-\lambda_{H,2}} \bigg ](\lambda_{H,2})^2 \ . 
\label{thkr}
\eeq
\end{widetext}

\subsection{Chromatic and Flow Polynomials}

Using Eq. (\ref{ptrel}), one readily verifies that for the case $x=1-q$, $y=0$,
the Tutte polynomial (\ref{thkr}) yields the chromatic polynomial for $H_{k,r}$
that we calculated in (Eq. (3.7) of) Ref. \cite{wa3}.

One can also discuss other special cases.  For the flow polynomial
$F(H_{k,r},q)$, we set $x=0$, $y=1-q$ and, with eq. (\ref{ftrel}), we have 
\beq
F(H_{k,r},q) = q^{-1}[(q-1)^r + (q-1)(-1)^r ] \ . 
\label{fhkr}
\eeq
This is independent of $k$, in accordance with the general result 
(\ref{flowvertexinflation}).  If $G$ is a planar graph and
$G^*$ is its planar dual, then the flow and chromatic polynomials satisfy the
relation 
\beq
F(G,q) = q^{-1}P(G^*,q) \ . 
\label{fprel}
\eeq
We observe that $F(H_{k,r},q) = q^{-1}P(C_r,q)$, in accord with
(\ref{flowvertexinflation}) and the fact that $H_{2,r}=(C_r)^*$ (cf. Eq. 
(\ref{hk2rdualcr})).

\subsection{Reliability Polynomial}

A communication network, such as the internet, can be represented by a graph,
with the vertices of the graph representing the nodes of the network and the
edges of the graph representing the communication links between these nodes.
In realistic networks, both the nodes and the links between them are imperfect,
and fail to operate.  One common measure of the reliability of the network is
the probability that there is a working communications route between any node
and any other node.  This is the all-terminal reliability function.  This is
commonly modeled by a simplification in which one assumes that each node and
link are operating with respective probabilities $p_{node}$ and $p_{link}$. 
As probabilities, $p_{node}$ and $p_{link}$ lie in the interval
[0,1]. The dependence of the all-terminal reliability function
$R_{tot}(G,p_{node},p_{link})$ on $p_{node}$ is an overall factor of
$(p_{node})^n$; i.e., $R_{tot}(G,p_{node},p_{link}) = 
(p_{node})^n R(G,p_{link})$.  Thus, the difficult part of the calculation of 
$R_{tot}(G,p_{node},p_{link})$ is the determination of $R(G,p_{link})$.  For
notational brevity, we set $p_{link} \equiv p$. 
The function $R(G,p)$ is given by 
\beq
R(G,p) = \sum_{\tilde G \subseteq G} p^{e(\tilde G)} \,(1-p)^{e(G)-e(\tilde G)}
\ , 
\label{rgen}
\eeq
where $\tilde G$ is a connected spanning subgraph of $G$.  Clearly,
$R(G,p)$ is a monotonically increasing function of $p \in [0,1]$ with
the boundary values $R(G,0)=0$ and $R(G,1)=1$.  $R(G,p)$ can be related to a
special case of the Tutte polynomial, evaluated with $x=1$ 
(guaranteeing that $\tilde G$ is a connected spanning subgraph of $G$) 
and $y=1/(1-p)$.  This relation is 
\beq
R(G,p)=p^{n-1}(1-p)^{e(G)+1-n} \, T(G,1,\frac{1}{1-p}) \ .
\label{rtrel}
\eeq
As in our previous calculations for lattice strips \cite{r}, we can thus 
obtain reliability polynomials as special cases of Tutte polynomials. 
Using our calculation in Eq. (\ref{thkr}) of $T(H_{k,r},x,y)$ with $x=1$ and
$y=1/(1-p)$ in Eq. (\ref{rtrel}), we have calculated $R(H_{k,r},p)$.  For 
$r=2$, we have 
\beq
R(H_{k,2},p) = R(C_{2k-2},p) = p^{2k-3}[p+2(k-1)(1-p)] \ ,
\label{rhk2}
\eeq
in accord with Eq. (\ref{hkr2}).  For $k=2$, i.e., the case of no 
homeomorphic expansion, we find, in accord with Eq. (\ref{hk2rdualcr}), that
\beq
R(H_{2,r},p) = R((C_r)^*,p) = 1-(1-p)^r \ . 
\label{rfatlinkr}
\eeq
In general, we find that for fixed $p \in (0,1)$ and fixed $r$, $R(H_{k,r},p)$
is a monotonically decreasing function of $k$.  This can be interpreted as a 
consequence of the fact that as $k$ increases, the girth $g(H_{k,r})$
increases (cf. Eq. (\ref{girth_hkr})), and hence there is a greater likelihood
that one of the communication links along the minimum-distance path 
and other paths between two nodes is not operating.

In contrast, for fixed $p \in (0,1)$ and fixed $k \ge 3$, we find a variety of 
behaviors for $R(H_{k,r},p)$ as a function of $r$. To illustrate this, we 
take $k=3$ and compare a few pairs of values $(r,r')=(r,2)$.  We calculate
\beq
R(H_{3,2},p) = p^3(4-3p) \ , 
\label{rhk3r2}
\eeq
\beq
R(H_{3,3},p) = p^4(7p^2-18p+12) \ , 
\label{rhk3r3}
\eeq
\beq
R(H_{3,4},p) = p^5(4-3p)(5p^2-12p+8) \ . 
\label{rhk3r4}
\eeq
and
\beq
R(H_{3,5},p) = p^6(31p^4-150p^3+280p^2-240p+80)
\label{rhk3r5}
\eeq
Hence, for the pairs $(r,r')=(r,2)$, 
\beq
R(H_{3,3},p)-R(H_{3,2},p) = p^3(1-p)^2(7p-4) \ , 
\label{rhk3r3_minus_rhk3r2}
\eeq
\beq
R(H_{3,4},p)-R(H_{3,2},p) = p^3(1-p)^2(4-3p)(5p^2-2p-1) \ , 
\label{rhk3r4_minus_rhk3r2}
\eeq
and
\beq
R(H_{3,5},p)-R(H_{3,2},p) = p^3(1-p)^2(31p^5-88p^4+73p^3-6p^2-5p-4)
\label{rhk3r5_minus_rhk3r2}
\eeq
As is evident from the difference in Eq. (\ref{rhk3r3_minus_rhk3r2}), 
\beq
R(H_{3,3},p) > R(H_{3,2},p) \quad {\rm if} \quad 1 > p > \frac{4}{7} 
\simeq 0.571 \ , 
\label{rhk3r3_versus_rhk3r2_largep}
\eeq
while 
\beq
R(H_{3,3},p) < R(H_{3,2},p) \quad {\rm if} \quad 0 < p < \frac{4}{7} \ . 
\label{rhk3r3_versus_rhk3r2_smallp}
\eeq
Similarly, Eq. (\ref{rhk3r4_minus_rhk3r2}) shows that 
\beq
R(H_{3,4},p) > R(H_{3,2},p) \quad {\rm if} \quad 1 > p >
\frac{1+\sqrt{6}}{5} \simeq 0.690 \ , 
\label{rhk3r4_versus_rhk3r2_largep}
\eeq
while 
\beq
R(H_{3,4},p) < R(H_{3,2},p) \quad {\rm if} \quad 0 < p < \frac{1+\sqrt{6}}{5}
 \ . 
\label{rhk3r4_versus_rhk3r2_smallp}
\eeq
Similarly, $R(H_{3,5},p) > R(H_{3,2},p)$ for $1 > p > 0.8418$, and
$R(H_{3,5},p) < R(H_{3,2},p)$ for $0.8418 > p > 0$ (where the crossover value
of $p$ is a root of the quintic in Eq. (\ref{rhk3r5_minus_rhk3r2}) quoted to
four significant figures). Note how, for a fixed value of $k$, the value of $p$
beyond which $R(H_{3,r},p)$ is greater than $R(H_{3,2},p)$ increases with $r$,
from 0.571 for $r=3$ to 0.690 for $r=4$ to 0.842 for $r=5$ (to three figures
accuracy), and so forth for higher values of $r$. We observe similar behavior
for $R(H_{k,r},p)-R(H_{k,2},p)$ as a function of $r$ for higher values of $k$.

However, we also have 
\beq
R(H_{3,4},p)-R(H_{3,3},p) = -p^4(1-p)^2(15p^2-26p+12) \ , 
\label{rhk3r4_minus_rhk3r3}
\eeq
and
\beq
R(H_{3,5},p)-R(H_{3,4},p) = -p^5(1-p)^2(-31p^3+88p^2-88p+32) \ .
\label{rhkr3r5_minus_rhk3r4}
\eeq
These comparisons provide a contrasting type of behavior, since 
\beq
R(H_{3,5},p) < R(H_{3,4},p) < R(H_{3,3},p) \quad \forall \ p \in (0,1) \ . 
\label{rhk3r4_versus_rhk3r3_smallp}
\eeq
and so forth with $R(H_{3,r},p)$ for larger values of $r$. Thus, for these
cases, for all $p \in (0,1)$, $R(H_{3,r},p)$ is a monotonically decreasing
function of $r$ as $r$ increases above 3.

\subsection{Some Graphical Quantities} 

We next discuss special valuations of $T(H_{k,r},x,y)$ that yield quantities of
graph-theoretic interest. For $x=1$, the following result is convenient:
\beqs
& & T(H_{k,r},1,y) = (k-1)^{r-2}(2k+y-3) \cr\cr
& + & \bigg [ \frac{(k+y-2)^{r-2}-(k-1)^{r-2}}{y-1} \bigg ](k+y-2)^2 \ . \cr\cr
& & 
\label{thkrx1y}
\eeqs
In addition to 
\beq
N_{SSG}(H_{k,r})=2^{(k-1)r} \ , 
\label{ssghkr}
\eeq
we compute 
\beq
N_{ST}(H_{k,r}) = r(k-1)^{r-1} \ , 
\label{thkrx1y1}
\eeq
\beq
N_{SF}(H_{k,r}) = (2^{k-1}-1)^{r-2}\Big [
  2^{2(k-1)}-1+(r-2)(2^{k-1}-1) \Big ] \ , 
\label{thkrx2y1}
\eeq
and
\beq
N_{CSSG}(H_{k,r}) = k^r-(k-1)^r \ . 
\label{thkrx1y2}
\eeq

Since these are two-parameter families of graphs, one can consider the limits 
(i) \ $r \to \infty$ with fixed finite $k$, and 
(ii) \ $k \to \infty$ with fixed finite $r$. 
We calculate the growth constants for each of these.  The cases of interest
here are those with $k \ge 3$, i.e., those with homeomorphic expansion.  For
these we find that for the limit (i) , 
\beq
(i): \ w_{_{SSG}} = 2^{\frac{k-1}{k-2}}  \ , 
\label{whkr_22_limr} 
\eeq
\beq
(i): \ w_{_{ST}} = (k-1)^{\frac{1}{k-2}} \ , 
\label{whkr_11_limr}
\eeq
\beq
(i): \ w_{_{SF}} = (2^{k-1}-1)^{\frac{1}{k-2}} \ , 
\label{whkr_21_limr}
\eeq
and
\beq
(i): \ w_{_{CSSG}} = k^{\frac{1}{k-2}} \ . 
\label{whkr_12_limr}
\eeq
For the limit (ii), we calculate
\beq
(ii): \ w_{_{SSG}} = w_{_{SF}} = 2
\label{whkr_22_21_limk}
\eeq
and
\beq
(ii): \ w_{_{ST}} = w_{_{CSSG}} = 1 \ . 
\label{whkr_11_12_limk}
\eeq

\section{Conclusions}

In conclusion, in this paper have have derived exact relations between the
Potts model partition function, or equivalently, the Tutte polynomial, for a
graph $G$ and for a graph $\tilde G$ obtained from $G$ by edge or vertex
inflation.  An analysis was given of some physical effects of uniform
$\ell$-fold edge and vertex inflations.  We have presented exact calculations
of the Tutte polynomials for free, cyclic, and M\"obius ladder families of
graphs $S_{k,m}$, $L_{k,m}$, and $M_{k,m}$ of length $m$, with $\ell=k-2$
vertex inflation on all longitudinal edges. We have given similar results for
the Tutte polynomial of the family $H_{k,r}$ of hammock graphs.  Our present
results generalize our previous calculations in Refs. \cite{wa3,pg,a}. As one
application, we have calculated reliability polynomials for the hammock
graphs and analyzed their properties as a function of $k$ and $r$.  In
addition, we have used our calculations to compute the number of spanning
trees, spanning forests, and connected spanning subgraphs on these families and
to determine their asymptotic behavior as the number of vertices goes to
infinity.

\begin{acknowledgments}

This research was partially supported by the grant NSF-PHY-06-53342.

\end{acknowledgments}


\begin{thebibliography}{99}

\bibitem{wurev}
F. Y. Wu, Rev. Mod. Phys. {\bf 54}, 235 (1982).

\bibitem{tutte}
W. T. Tutte, {\it Graph Theory}, in G. C. Rota, ed., 
{\it Encyclopedia of Mathematics and its Applications}, 
vol. 21 (Addison-Wesley, New York, 1984). 

\bibitem{biggs}
N. Biggs, {\it Algebraic Graph Theory} (Cambridge Univ. Press, Cambridge,
1993).

\bibitem{boll}
B. Bollob\'as, {\it Modern Graph Theory} (Springer, New York, 1998).

\bibitem{chowwu}
Y. Chow and F. Y. Wu, Phys. Rev. B {\bf 36}, 285 (1987).

\bibitem{w}
R. Shrock and S.-H. Tsai, Phys. Rev. E {\bf 55}, 5165 (1997).

\bibitem{wa3}
R. Shrock and S.-H. Tsai, J. Phys. A {\bf 31}, 9641 (1998). 

\bibitem{wa2}
R. Shrock and S.-H. Tsai, Physica {\bf A265}, 186 (1999). 

\bibitem{pg}
R. Shrock and S.-H. Tsai, J. Phys. A Letts. {\bf 32}, L195 (1999). 

\bibitem{a}
R. Shrock, Physica A {\bf 283}, 388 (2000).

\bibitem{welsh}
%
D. J. A. Welsh, {\it Complexity: Knots, Colourings and
Counting}, London Math. Soc. Lecture Notes vol. {\bf 186} (Cambridge
Univ. Press, Cambridge, 1993).

\bibitem{jemrev}
L. Beaudin, J. Ellis-Monaghan, G. Pangborn, and R. Shrock, Discrete Math. 
{\bf 310}, 2037 (2010).

\bibitem{fk}
C. M. Fortuin and P. W. Kasteleyn, Physica {\bf 57}, 536 (1972). 

\bibitem{evd}
%
Note that with our labelling conventions, an $\ell$-fold edge inflation with
$\ell=1$ means to replace that edge with an equivalent single edge, which
leaves the graph unchanged; the choices $\ell \ge 2$ change the graph.  In
contrast, for an $\ell$-fold vertex inflations, the choice $\ell=0$ leaves the
graph unchanged, while the choices $\ell \ge 1$ change the graph.

\bibitem{wz84}
E. G. Whitehead and L. C. Zhao, J. Graph Theory {\bf 8}, 355 (1984).

\bibitem{hs}
R. Shrock and S.-H. Tsai, Physica A {\bf 259}, 315 (1998).

\bibitem{rw99}
R. C. Read and E. G. Whitehead, Discrete Math. {\bf 204}, 337 (1999);
{\it ibid.} {\bf 243}, 267 (2002); {\it ibid.} {\bf 308}, 1826 (2008). 

\bibitem{traldi00}
L. Traldi, Discrete Math. {\bf 220}, 291 (2000); {\it ibid.} {\bf 248}, 
279 (2002). 

\bibitem{sokal04}
A. Sokal, Combin. Probab. Comput. {\bf 13}, 221 (2004). 

\bibitem{dktbook}
Dong, F. M., Koh, K. M., Teo, K. L.: {\it Chromatic Polynomials and
Chromaticity of Graphs} (World Scientific, Singapore, 2005).

\bibitem{kdec}
R. Shrock and Y. Xu, arXiv:1101.0852. 

\bibitem{syozi}
%
An early review is I. Syozi, in C. Domb and M. S. Green, eds., {\it Phase
Transitions and Critical Phenomena} (Academic Press, New York, 1972), p. 269.

\bibitem{onsager44}
L. Onsager, Phys. Rev. {\bf 65}, 117 (1944). 

\bibitem{spin_glasses}
%
D. Sherrington and S. Kirkpatrick, Phys. Rev. Lett. {\bf 35}, 1792 (1975);
D. Elderfield and D. Sherrington, J. Phys. C {\bf 16}, L497 (1983);
H. Nishimori and M. J. Stephen, Phys. Rev. B {\bf 27}, 5644 (1983);
M. M\'ezard, G. Parisi, and M. A. Virasoro, {\it Spin Glass Theory and 
Beyond} (World Scientific, Singapore, 1987).


\bibitem{bbound}
N. L. Biggs, Bull. London Math. Soc. {\bf 9}, 54 (1977). 

\bibitem{wn}
R. Shrock and S.-H. Tsai, Phys. Rev. E {\bf 55}, 6791 (1997); {\it ibid.}
{\bf 56}, 2733 (1997); {\it ibid.} {\bf 56}, 4111 (1997). 


\bibitem{stephenson}
J. Stephenson, J. Math. Phys.  {\bf 5}, 1009 (1964); {\it ibid.}, 
{\bf 11}, 413 (1970). 

\bibitem{rg}
%
Early reviews include M. E. Fisher, Rev. Mod. Phys. {\bf 46}, 597 (1974) and
C. Domb and M. S. Green, eds., {\it Phase Transitions and Critical Phenomena}
(Academic Press, New York, 1976), vol. 6.

\bibitem{eta_ising}
L. P. Kadanoff, Nuovo Cim. {\bf 44B}, 276 (1966); 
T. T. Wu, Phys. Rev. {\bf 149}, 380 (1966). 

\bibitem{pm}
R. Shrock, Phys. Lett. A {\bf 261}, 57 (1999).

\bibitem{cf}
S.-C. Chang and R. Shrock, Physica A {\bf 296}, 131 (2001).

\bibitem{fp}
S.-C. Chang and R. Shrock, J. Stat. Phys., {\bf 112}, 815 (2003). 

\bibitem{complexnets}
%
Recent reviews include M. Newman, A.-L. Barab\'asi, and D. J. Watts, {\it
The Structure and Dynamics of Networks} (Princeton Univ. Press, Princeton,
2006) and S. N. Dorogovstev, A. V. Goltsev, and
J. F. F. Mendes, Rev. Mod. Phys. {\bf 80}, 1275 (2008);

\bibitem{r}
S.-C. Chang and R. Shrock, J. Stat. Physics {\bf 112}, 1019 (2003). 

\end{thebibliography}
\end{document}